\documentclass[lettersize,journal]{IEEEtran}
\usepackage{amsmath,amsfonts}
\usepackage{algorithmic}
\usepackage{algorithm}
\usepackage{array}
\usepackage{calc}
\usepackage[caption=false,font=normalsize,labelfont=sf,textfont=sf]{subfig}
\usepackage{textcomp}
\usepackage{stfloats}
\usepackage{multirow}
\usepackage{url}
\usepackage{verbatim}
\usepackage{amsmath}
\usepackage{mathtools}
\usepackage{relsize}
\usepackage{graphicx}
\usepackage{cite}
\usepackage[none]{hyphenat}
\usepackage{bm}
\usepackage[dvipsnames]{xcolor}
\begin{document}
\title{{Low-complexity Samples versus Symbols-based Neural Network Receiver for Channel Equalization}}
\author{Yevhenii~Osadchuk,~\IEEEmembership{Student Member,~IEEE,}
Ognjen~Jovanovic,~\IEEEmembership{Member,~IEEE,}
Stenio~M.~Ranzini,
Roman~Dischler,
Vahid~Aref,
Darko~Zibar,
and~Francesco~Da~Ros~\IEEEmembership{Senior Member,~IEEE,}
\thanks{Manuscript received June 30, 2023; revised August 16, 2023.}
\thanks{Yevhenii~Osadchuk, Ognjen~Jovanovic, Darko~Zibar~and~Francesco~Da~Ros are with DTU Electro, Technical University of Denmark, 2800 Kgs. Lyngby, Denmark. (e-mails: yevos@dtu.dk; dazi@dtu.dk; fdro@dtu.dk)}
\thanks{Stenio~M.~Ranzini was with DTU Electro, Technical University of Denmark, 2800 Kgs. Lyngby, Denmark. and now is with Infinera, Nuremberg. (e-mail: smara@dtu.dk)} 
\thanks{Roman~Dischler~and~Vahid~Aref are with Nokia Bell Labs, Stuttgart, Germany. (e-mails: roman.dischler@nokia.com; vahid.aref@nokia.com)}}
\markboth{Journal of \LaTeX\ Class Files,~Vol.~14, No.~8, August~2023}%
{Shell \MakeLowercase{\textit{et al.}}: A Sample Article Using IEEEtran.cls for IEEE Journals}
\maketitle

\begin{abstract}
Low-complexity neural networks (NNs) have successfully been applied for digital signal processing (DSP) in short-reach intensity-modulated directly detected optical links, where chromatic dispersion-induced impairments significantly limit the transmission distance. The NN-based equalizers are usually optimized independently from other DSP components, such as matched filtering. This approach may result in lower equalization performance. Alternatively, optimizing a NN equalizer to perform functionalities of multiple DSP blocks may increase transmission reach while keeping the complexity low. In this work, we propose a low-complexity NN that performs samples-to-symbol equalization, meaning that the NN-based equalizer includes match filtering and downsampling. We compare it to a samples-to-sample equalization approach followed by match filtering and downsampling in terms of performance and computational complexity. Both approaches are evaluated using three different types of NNs combined with optical preprocessing. We numerically and experimentally show that the proposed samples-to-symbol equalization approach applied for 32 GBd on-off keying (OOK) signals outperforms the samples-domain alternative keeping the computational complexity low. Additionally, the different types of NN-based equalizers are compared in terms of performance with respect to computational complexity.
\end{abstract}

\section{Introduction}
\IEEEPARstart{I}NTENSITY-modulated and directly detected (IM/DD) transceivers have been widely implemented due to their low cost, simplicity, and low footprint, making them highly suitable for applications requiring a large number of transceivers, such as short-reach interconnects~\cite{KARANOV202265, 9122234}. However, the transmission reach of IM/DD links is limited by the intersymbol interference (ISI) effect induced by the linear chromatic dispersion (CD) accumulated during fiber propagation and the nonlinear square-law photo-detector (PD). To tackle the nonlinear impairments and extend transmission reach the application of a nonlinear equalizer is a requirement for IM/DD receivers operating in the C-band. Although various equalization techniques have been proposed, the search for high-performance and low-complexity digital signal processing (DSP) remains ongoing~\cite{9751344}.

Due to the rapid development of machine learning computing frameworks, various types of neural network (NN)-based equalizers have been proposed as promising solutions in short-reach fiber transmission outperforming traditional equalization techniques~\cite{9751344, 7767495, 9560044, Xu:19, 8535153}. The NNs in the shape of feedforward NNs (FNN)~\cite{9560044}, recurrent NNs (RNN)~\cite{Xu:19, 8535428, 9726927}, and convolutional NNs (CNN)~\cite{8386096} show higher equalization performance compared to conventional Volterra~\cite{Schdler2021SoftDemappingFS} or feedforward equalizer (FFE)~\cite{8535428}, effectively addressing nonlinearities in IM/DD links. However, the NN-based equalizers that have high equalization capabilities are often highly complex~\cite{9751344}. Therefore, there is a need to develop a low-complexity NN-based equalizer that can effectively tackle nonlinear channel impairments.

Recently, an equalization proposal has emerged that combines optical pre-processing with digital NNs to divide the complexity between optical and electrical domains~\cite{9314705, 10144303, Sozos2022HighspeedPN}. It was experimentally shown that reservoir computing can compensate for CD when applied for a sequence of samples of time domain signal. Such equalizers are usually optimized taking into account receiver-side match filters and downsampling. However, optimizing such equalizers independently from the rest of the DSP, such as match filtering, can lead to a lower equalization accuracy~\cite{9018053, 9560044}. To address this, an alternative approach involves embedding multiple DSP blocks into a single FNN-based equalizer and optimizing it as a unified entity~\cite{9975476}. This approach has the potential to enable NNs to fully perform symbols recovery and avoid additional postprocessing complexity in the shape of matched filtering.
\begin{figure*}[ht!]
    \centering
    \includegraphics[width=\textwidth]{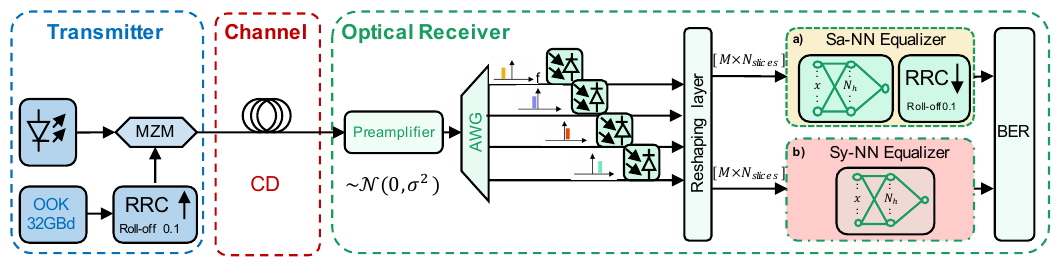}
    \caption{Communication setup under investigation. a) Sa-NN equalizer, and b) Sy-NN equalizer.}
    \label{fig:setup1}
\end{figure*}

In this work, we extend an investigation of FNN equalizers from~\cite{9975476} by showing different NN's performance from the perspective of the impact of computational complexity. We propose using a samples-to-symbol NN-based equalizer that performs the task of downsampling,  match filtering of the pulse shape, and equalization together as one NN receiver block.  We compare this approach to the samples-to-sample NN-based equalizer (the idea is to reconstruct the information from the sliced spectrum together with equalization), followed by the matched filter of the pulse shaping and downsampling. Additionally, we compare the performance of two approaches in the shape of different NN types such as FNN, RNN, and CNN. We show that the samples-to-symbol approach significantly outperforms the samples-to-sample equalization in terms of bit error ratio (BER) performance for all types of NN. Furthermore, we investigate the complexity of the proposed NN and show that it provides low BER performance while keeping low computational complexity suitable for digital equalizer implementation. Finally, we numerically and experimentally show that samples-to-symbol FNN requires fewer or the same number of multiplications per symbol than the rest of the investigated equalizers, extending the transmission distance for complexity-constrained applications.

The structure of the paper is as follows: Section II outlines the numerical and experimental setup used for short-reach transmission. Section III provides a description of the various NN-based equalizers that were examined. In Section IV, we explain the method for calculating the computational complexity of all the NN-based equalizers discussed. The performance and complexity of the NN equalizers are compared and highlighted in Section V. Finally, we summarize our findings in the conclusion section.

\section{Short-Reach System Under Investigation}
\subsection{Numerical Setup}
First, we describe the numerical setup used to investigate NN-based equalizers. The simulation setup is shown in Fig.~\ref{fig:setup1}. At the transmitter, a pseudorandom sequence of $2^{21}$ bits is generated using a Mersenne Twister generator, upsampled by 8 samples per symbol (sps), and shaped by root-raised-cosine RRC filter ($\alpha=0.1$) to generate a 32-GBd OOK signal. Then, the electrical signal is encoded in the optical domain using a Mach-Zehnder modulator (MZM) and input into the communication channel. As our goal is to investigate the impact of CD on the signal, the standard single-mode fiber (SSMF) transmission is modeled only with CD ($D=16.4$~ps/nm/km). The receiver pre-amplifier noise is modeled as an additive white Gaussian noise (AWGN) source with a tunable variance ($\sigma^2$) which allows adjusting the signal-to-noise ratio (SNR) of the received signal. After amplification, the signal`s spectrum is divided into $N_{slices}=4$ equal slices by an arbitrary waveguide grating (AWG) to reduce the power fading effect, as proposed in~\cite{9018053}. Numerically, the AWG can be designed as a fixed number of second-order Gaussian filters with a 3-dB bandwidth of 16 GHz to slice the signal into $N_{slices}=4$ number of equal slices of 8 GHz that is fixed throughout the rest of the work. More details of signal slicing can be found in~\cite{9314705, 9018053}. After that, each slice of the total signal is detected by the separate photodetector (PD) in a square-law fashion. Before feeding the signal into the NN equalizer, the sequence is reshaped to create a serial structure via a reshaping layer (Fig.~\ref{fig:setup1}). Then, the signal sequence is fed into two types of NN-based equalization: samples-to-sample (Sa-NN) and samples-to-symbol (Sy-NN). The reshaping procedure and the details of the input-output structure for both Sa-NN and Sy-NN are described in Section \ref{sec3}. Finally, the BER is estimated through error counting as a function of SNR and distance.

\subsection{Experimental Setup}
To validate the numerical results, the experimental scenario used in this work is adopted from~\cite{9314705}. In the experiment, the signal is shaped with RRC ($\alpha=0.1$) and upsampled to 2 sps. The signal is then resampled to match the sampling rate of the digital-to-analog converter at 88 GSa/s. Next, the signal is modulated using MZM with the bias adjusted to the quadrature point. The optical signal is then transmitted through 74 km of a SMF. At the receiver, the signal is first amplified using an erbium-doped fiber amplifier (EDFA) and then filtered by a wavelength selective switch (WSS) modeling the AWG. The filters are configured as second-order Gaussian filters with a 3-dB bandwidth of 16 GHz, as for the numerical analysis. Subsequently, the signal is independently detected by four PDs, each having the same bandwidth of 40 GHz. At the receiver, the signal is digitally resampled using a real-time oscilloscope operating at a sampling rate of 80 Gsa/s and electrical bandwidth of 33 GHz. After the detection, the signal is post-processed offline with a low-pass filter, followed by equalization as in numerical simulations. Then, the symbols' hard decision is made and the BER is calculated. The structure of the equalization approaches is defined in Section \ref{sec3}.

\section{Neural Networks-based Receiver Design}\label{sec3}
In this section, we describe the Sa-NN and the Sy-NN equalizer architectures. First, both Sy-NN and Sa-NN equalizers are applied to the sliced time domain signal right after the PD. Both approaches employ a simple NN architecture consisting of one input layer, one hidden layer, and one output layer. The input to the NN is defined with a memory $M$ and a number of features $N_{slices}$ of optical slices of the same signal that correspond to $M \cdot N_{slices}$ total input. Each sample in memory $M$ is sliced into $N_{slices}=4$ and it is fixed throughout the rest of the work. The size of the memory $M$ is determined by the formula:
\begin{gather}\label{eq:1}
M = 2L + q,
\end{gather}
where $L=K\cdot sps$ represents the $K$ number of previous and future symbols upsampled by $sps$, while $q$ represents the current equalized unit. For Sa-NN $q=1$, because it is aimed to process 1 input sample and provide 1 sample at the output. For Sy-NN, however, the $q=sps$ as is aimed to equalize all the samples that belong to 1 symbol at the output. The serial structure of the total input sequence will be defined in more detail in the following paragraph. It is important to highlight that, for numerical simulations, the Sa-NN equalizers are applied for the signal with $sps=8$. Reducing the number of sps significantly degrades equalization performance, therefore keeping $sps=8$ is essential for all Sa-NN equalizers. The performance of Sy-NN instead, remains stable for both  $sps=8$ and $sps=2$ samples per equalized symbol. Therefore, in this work, for Sa-NN - the signal is upsampled at 8 sps, while for Sy-NN $sps=2$ is used. However, in the experimental scenario, $sps=2$ were used for both Sa-NN and Sy-NN cases due to the limited sampling rate of the ADCs. Due to the application of time delay memory the number of sps plays an important role in defining computational complexity as will be discussed in the next sections.

Due to the equalization of a single sample in Sa-NN and a single symbol in Sy-NN, the input sequences to both NN equalizers are different in size. If we define\break the $\mathbf{x}^{r}_{k}=[x^{(1)}_{k}, x^{(2)}_{k}, x^{(3)}_{k}, x^{(4)}_{k}]$ to be a set of 4 slices of a $k$-th sample with $L$ number of previous or forward samples, the input to Sa-NN can be defined as:
\begin{gather}\label{eq:2}
\begin{split}
\mathbf{x}^{Sa} = \Bigr[&\mathbf{x}^{r}_{k-L}, 
\mathbf{x}^{r}_{k-(L-1)},..., 
\underbrace{\mathbf{x}^{r}_{k}}_{\text{\scriptsize sample}},..., 
\mathbf{x}^{r}_{k+(L-1)},
\mathbf{x}^{r}_{k+L}\Bigr]
\end{split}
\end{gather}

For the input of a Sy-NN equalizer, first we define\break $\mathbf{x}^{s}_t = [\mathbf{x}^{r}_{k}, \mathbf{x}^{r}_{k+1}, \mathbf{x}^{r}_{k+2},...,\mathbf{x}^{r}_{k+sps}]$ with $t=\lfloor k/sps\rfloor$, as a sequence of $sps$ samples that correspond to one input symbol. 
Then, the input to Sy-NN with $K$ previous or forward symbols can be defined as:
\begin{gather}\label{eq:3}
\begin{split}
\mathbf{x}^{Sy} = \Bigr[&
\mathbf{x}^{s}_{t-K}, 
\mathbf{x}^{s}_{t-(K-1)},...,
\underbrace{\mathbf{x}^{s}_{t}}_{\text{\scriptsize symbol}},...,
\mathbf{x}^{s}_{t+(K-1)}, 
\mathbf{x}^{s}_{t+K}\Bigr]. 
\end{split}
\end{gather}

For both Sa-NN and Sy-NN, the hidden layer is defined with $N_{h}$ hidden units. At the output, a single neuron represents 1 equalized sample for Sa-NN or 1 equalized symbol for Sy-NN equalizers. The activation function and other hyperparameters are determined through Bayesian optimization (BO)~\cite{9280329}. The output activation function $f_{out}$ was found $sigmoid$ for Sa-based and $linear$ for Sy-based NNs. The mini-batch size was found to be 1800 for Sa-based and 1000 data samples for Sy-based equalizers. The time domain signal is re-scaled before the input to the NN by using an optimized variance parameter $var$. The learning rate $l_{rate}$ is equal to $0.5\times 10^{-2}$ for Sa-NN and $1\times10^{-2}$ for Sy-NN. The optimized hyperparameters are summarized in Table ~\ref{tab:table1}. The training process involves using a regression approach with a mean squared error loss function on $2^{19}$ training symbols and $2^{16}$ testing symbols. Backpropagation with stochastic gradient descent was used for learning. The following subsections will describe the architecture of each NN in detail.
\subsection{Feedforward Neural Network}
\begin{figure}
    \centering
   \captionsetup{justification=centering,margin=2cm}
    \includegraphics[width=0.4\textwidth]{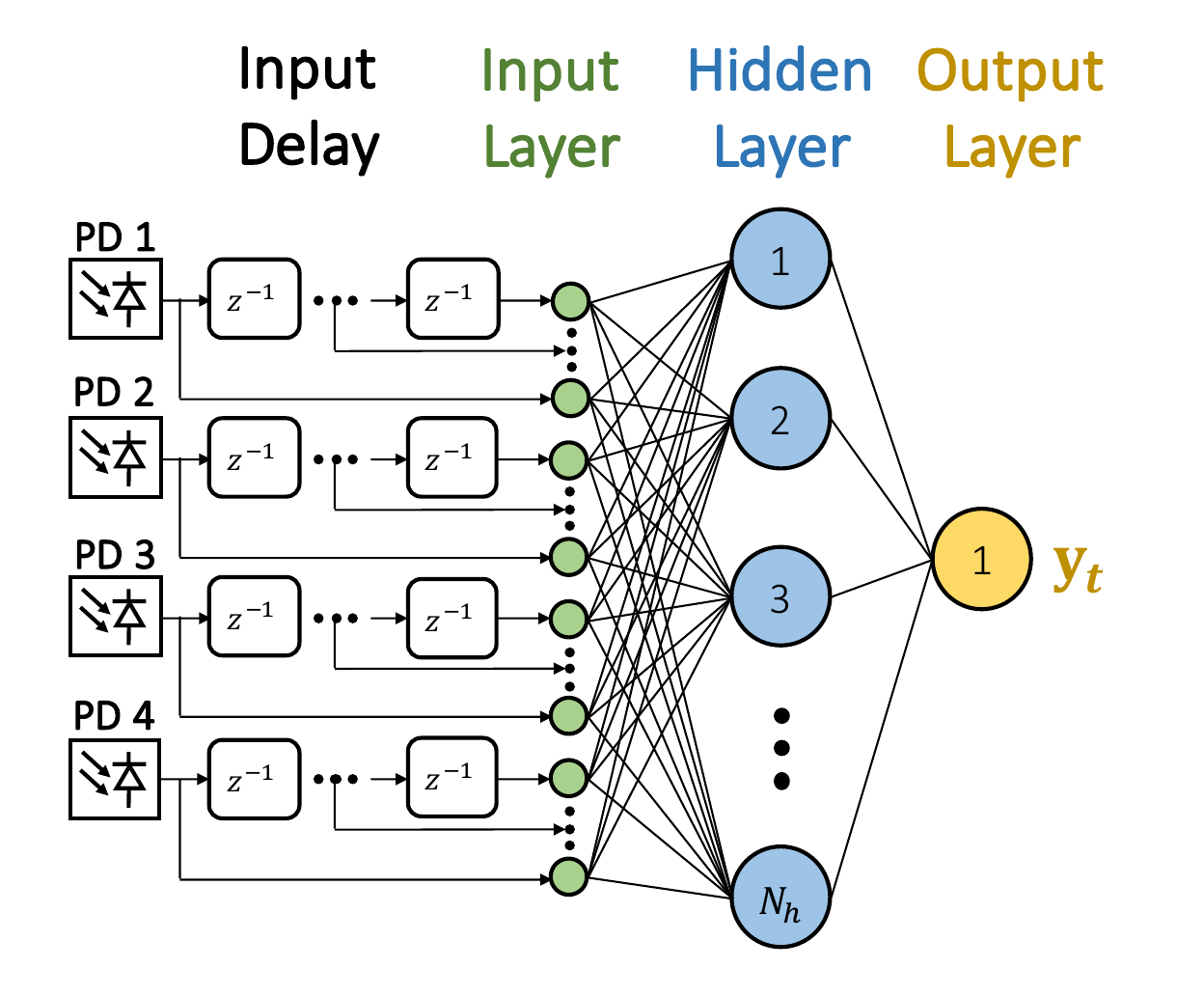}
    \caption{FNN-based equalizer structure with time-delayed input.}
    \label{fig:fnn}
\end{figure}

Feedforward neural network (FNN) is amongst the simplest options widely proposed for short-reach channel equalization, where densely connected structure helps effectively learn the memory-induced chromatic dispersion~\cite{Gaiarin:16,7767495,app9214675}. In this work, the input layer is designed with a time-delay window to support feedforward connectivity with additional short memory. The schematic of the FNN is shown in Fig.~\ref{fig:fnn}. To limit the complexity we consider the FNN with a single hidden layer and limit the number of hidden units to $N_h=10$. The hidden layer activation functions were found to be $sigmoid$ for Sa-FNN and $ReLU$ for Sy-FNN. The FNN with an input $x_t$ and single hidden layer for a single output $y_t$ recovery in a matrix form can be described by:
\begin{gather}
    y_t = f_{out}(f_{h}[x_{t}\cdot \bm{W}_h + b_h]\cdot \bm{W}_{out}),
\end{gather}
where $W_{h}$ and $W_{out}$ are the hidden and output weights matrices, and $f_h$ and $f_{out}$ are hidden and output activation functions. The rest of the hyperparameters were optimized using BO for a single scenario of 30 km transmission as an intermediate choice of distances considered in this work. It is worth noting that picking a specific transmission scenario for hyper-parameter optimization does not result in a significant change in equalization performance. The goal of BO in this case is more for choosing the best fitting NN hyperparameters rather than finding an extremely specific NN architecture that will provide the absolutely optimal performance. For example, optimization is used to find an appropriate hidden layer activation function out of the available functions that are suitable for this choice of NN application, instead of manually applying each function to find the best performance. Therefore, similar hyperparameter values were found when applying BO for other transmission distances.

\subsection{Gated Recurrent Unit Neural Network}
\begin{figure}[h]
    \centering
    \captionsetup{justification=centering,margin=2cm}
    \includegraphics[width=0.4\textwidth]{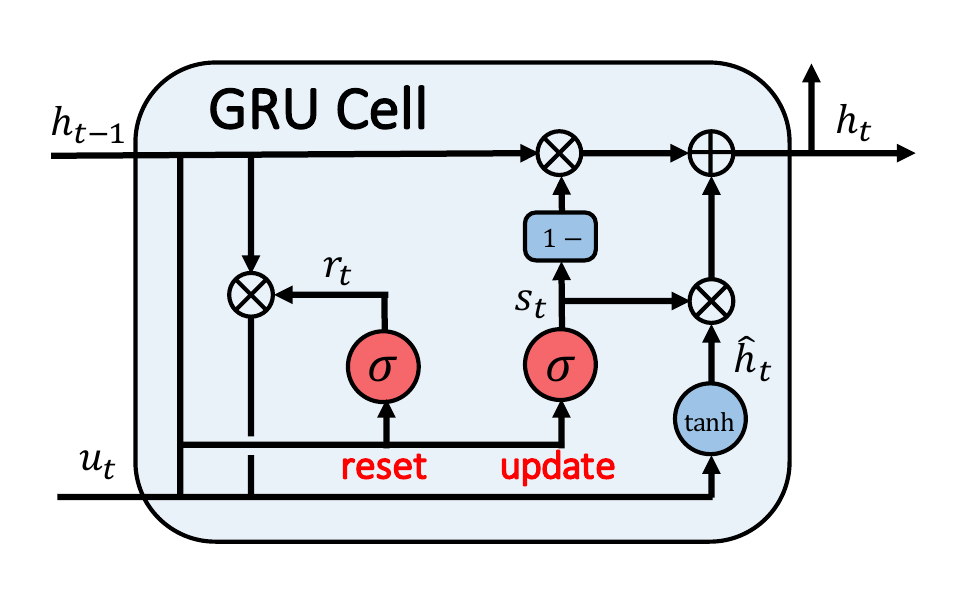}
    \caption{Schematics of gated architecture of a single GRU cell.}
    \label{fig:gru_cell}
\end{figure}

A gated recurrent unit is a special type of NN with recurrent connections that enhance the capabilities of traditional NNs by incorporating internal memory and gated architecture. By storing the hidden state at each time step, the GRUs are very suitable for the sequential data processing used within the DSP framework. The GRUs were successfully used in~\cite{9697352} to equalize 200 Gbps PAM-4 signal over 1 km transmission. The structure of a GRU cell is depicted in Fig.~\ref{fig:gru_cell}. The calculation of reset gate $\bm{r_{t}}$, update gate $\bm{s_{t}}$, a hidden state $\bm{h_{t}}$, and a candidate hidden state $\bm{\hat{h}_{t}}$ of a single hidden GRU cell with input $\bm{u_{t}}$ can be described as:
\begin{gather}
    \bm{r_{t}}=\sigma(\bm{W_{r}u_{t}}+\bm{U_{r}h_{t-1}}+\bm{b_{r}})\\
    \bm{s_{t}}=\sigma(\bm{W_{s}u_{t}}+\bm{U_{s}h_{t-1}}+\bm{b_{s}})\\
    \bm{\hat{h}_{t}}=tanh(\bm{W_{h}u_{t}}+\bm{r_{t}}\odot(\bm{U_{h}h_{t-1}}+\bm{b_{h}}))\\
    \bm{h_{t}}=(1-\bm{s_{t}})\odot \bm{h_{t-1}}-\bm{s_{t}}\odot \bm{\hat{h}_{t}} ,
\end{gather}
where $\bm{W_{r}}$, $\bm{U_{r}}$, $\bm{W_{s}}$, $\bm{U_{s}}$, $\bm{W_{h}}$, and $\bm{U_{h}}$ are corresponding weights matrices, and $\bm{b_{r}}$, $\bm{b_{z}}$, $\bm{b_{h}}$ are the biases. The $\sigma$ and $tanh$ are logistic sigmoid and hyperbolic tangent activation functions correspondingly. The $\odot$ is a Hadamard product. Following our idea with FNNs, here, we use a GRU network with a single layer and up to 10 hidden units to limit the maximum computational capacity. The activation function $f_h$ found by BO used for both Sy-GRU and Sa-GRU is $tanh$.
\subsection{Convolutional Neural Network}
\begin{figure}
    \centering
   \captionsetup{justification=centering,margin=2cm}
    \includegraphics[width=0.4\textwidth]{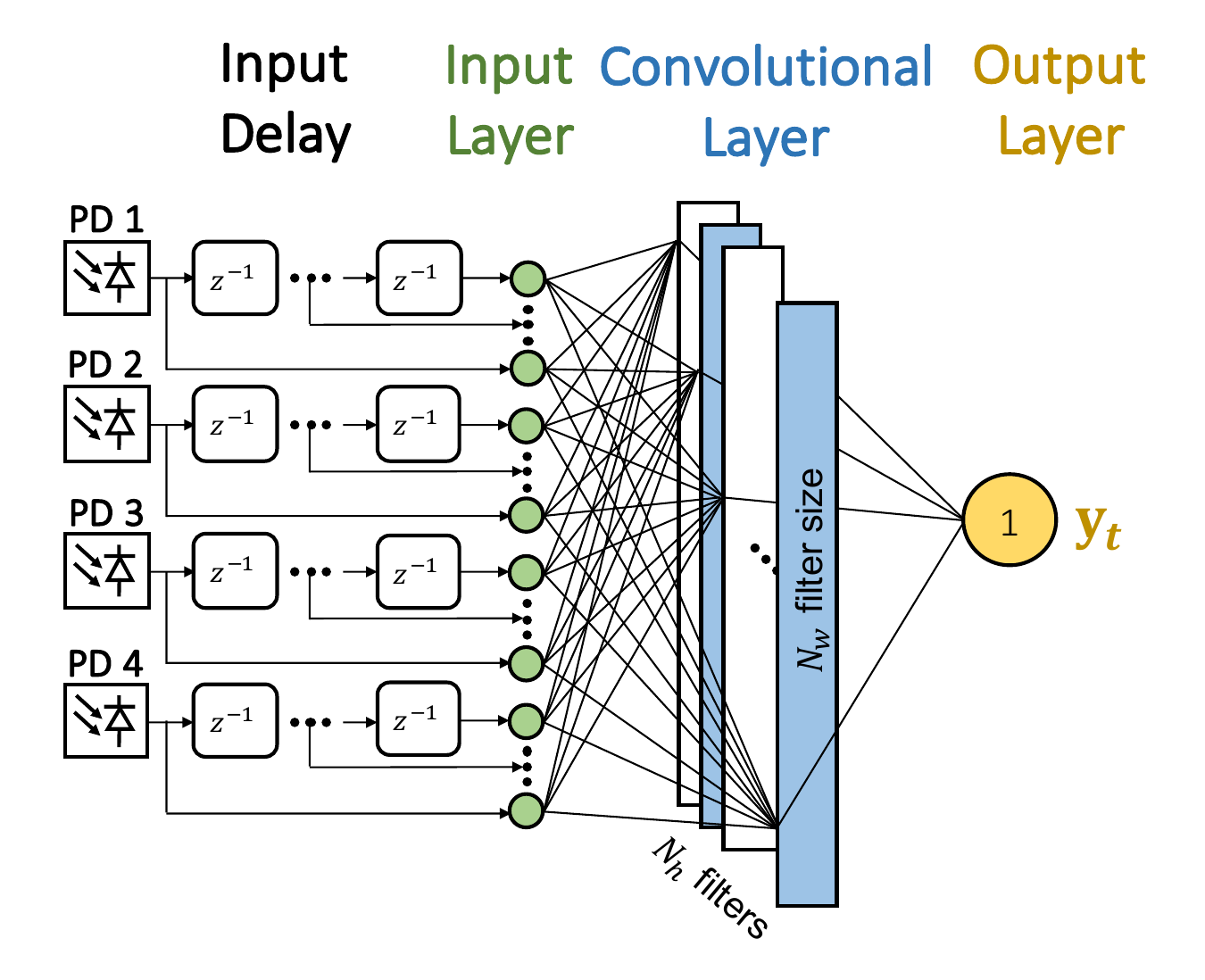}
    \caption{Architecture of CNN-based equalizer with time-delayed input.}
    \label{fig:cnn_pic}
\end{figure}

Convolutional Neural Network (CNN) is a powerful feed-forward filtering tool that was found effective in one-dimensional sequence processing, due to its ability to extract essential features from the input sequences. The CNN showed outstanding performance in equalizing 112 Gbps PAM-4 signal transmitted over 40 km SSMF~\cite{Chuang:18}. In~\cite{8385711}, the authors experimentally demonstrated the effectiveness of CNN applied for equalization of 56 Gbps PAM-4 IM/DD transmission over 25 km SSMF. In this work, we introduce the time delay to the input layer, which allows the CNN to have a short memory and filter multiple input features simultaneously. The architecture of CNN is shown in Fig.~\ref{fig:cnn_pic}. The convolution layer itself consists of several sliding filters $N_{h}$ with a size of $N_{w}$. Without downsampling by pooling and additional processing layers, the output of the convolution layer is equal to the size of  $N_{h}$. To simplify the convolution structure we fix the padding to 0, dilation to 1, and stride to 1. The input and output relationship can be defined as:
\begin{gather}
    y_{i}^{g}=f_h(\sum_{n=1}^{N_{slices}}\sum_{j=1}^{N_{w}}x_{i+j-1,n}\odot w_{j,n}^{g}+b_{j,n}^{g}),
\end{gather}
where $y_{i}^{g}$ defines the output feature map, for the $i-th$ input feature generated by filter $g$ of a convolution layer~\cite{Freire:21}. The $x$ is the input data vector, while $\mathbf{w}_j^g$ is the $j$-th kernel of a filter $g$ and $b_{j}^{g}$ is the bias. The $n$ index corresponds to the feature index in the range of 1 to $N_{slices}=4$, and $f_h$ is a nonlinear activation function. The output $y^{g}$ is then followed by the feedforward output layer. Optimized by BO, the number of filters $N_{h}$ is equal to 15, and the filter size $N_{w}$ is equal to 14 with the $sigmoid$ activation function.

The architectures of both Sa-NN and Sy-NN as well as their optimized hyperparameters are summarized in Table~\ref{tab:table1}.

\begin{table}[ht]
\caption{Hyperparameters found by Bayesian optimizer \label{tab:table1}}
\centering
\scalebox{0.89}{
\begin{tabular}{ |c||c|c|c||c|c|c| }
\hline
 & \multicolumn{3}{c||}{Sa-NN} & \multicolumn{3}{c|}{Sy-NN} \\
\hline
 & FNN & GRU & CNN & FNN & GRU & CNN \\
\hline
$sps$ & \multicolumn{3}{c||}{8} & \multicolumn{3}{c|}{2} \\
\hline
$K$ & \multicolumn{3}{c||}{3} & \multicolumn{3}{c|}{3} \\
\hline
$q$ & \multicolumn{3}{c||}{1} & \multicolumn{3}{c|}{$sps$} \\
\hline
$M$ & \multicolumn{3}{c||}{$49$} & \multicolumn{3}{c|}{$14$} \\
\hline
$N_{h}$ & $10$ & $10$ & $15$ & $10$ & $10$ & $15$ \\
\hline
$N_{w}$ & - & - & $49$ & - & - & $14$ \\
\hline
$f_{h}$ & $sigmoid$ & $tanh$ & $sigmoid$ & $ReLU$ & $tanh$ & $sigmoid$ \\
\hline
$f_{out}$ & \multicolumn{3}{c||}{$linear$} & \multicolumn{3}{c|}{$sigmoid$} \\
\hline
mini-batch & \multicolumn{3}{c||}{$1800$} & \multicolumn{3}{c|}{$1000$} \\
\hline
$var$ & \multicolumn{3}{c||}{$0.17$} & \multicolumn{3}{c|}{$0.69$} \\
\hline
$l_{rate}$ & \multicolumn{3}{c||}{$0.5\times 10^{-2}$} & \multicolumn{3}{c|}{$1\times10^{-2}$} \\
\hline
\end{tabular}
}
\end{table}


\section{Computational Complexity of NN-based equalizers}
The computational complexity of forward NN propagation represented by the number of multiplications per equalized symbol (RMPS) is one of the primary comparison methods in hardware channel equalization~\cite{9751344}. Multiplications consume most of the processing logic when dealing with float values with high decimal digits~\cite{Zot-mk}. In this work, we emphasize the focus on assessing the inference-phase computation complexity (CC) during the evaluation phase, rather than considering the training complexity of a NN, which is conducted offline during the calibration phase. Furthermore, our framework does not incorporate the computational complexity associated with nonlinear activation functions, as they often rely on approximation techniques rather than direct multiplicative calculations. Notably, in the traditional approach of lookup tables-based approximation, the application of such activation functions can be digitally implemented with significantly reduced computational requirements~\cite{716806}. The complexity of a feedforward NN-based equalizer~\cite{Freire:21} with input size (mini-batch, $M$, $N_{slices}$), single hidden layer with $N_{h}$ number of neurons, and a single output neuron $N_{out}$ can be defined as:
\begin{gather}
    CC_{FNN} = M N_{slices} N_{h} + N_{h} N_{out},
\end{gather}
The complexity of GRU NN is more complicated than the FNN structure due to the gated recurrent structure~\cite{Deligiannidis2021PerformanceAC}. The RMPS of a GRU equalizer is calculated as:
\begin{gather}
    \begin{split}
    CC_{GRU} = 3 (N_{slices} N_{h} + N_{h} N_{h}) M + N_{h} N_{out} M,
    \end{split}
\end{gather}
The complexity of a feed-forward CNN architecture composed of one single convolution layer~\cite{Freire:21} is defined as :
\begin{gather}
    \begin{split}
    CC_{CNN} = N_{slices} N_{h} N_{w} (M-N_{w}+1) \\ + (M-N_{w}+1) N_{h} N_{out}
    \end{split}
\end{gather}
where $N_{h}$ is the number of filters and $N_{w}$ is the filter size.

Finally, we also show the complexity of commonly used FFE as a benchmark comparison for the rest of the NNs. Important to mention, that the FFE can only be applied for the original signal without slicing with a single PD~\cite{9751344}. The complexity of an FFE can be calculated by considering the number of window taps $N_{taps}$ as:
\begin{gather}
    CC_{FFE} = N_{taps}+1
\end{gather}
The FFE is implemented with $N_{taps} = 11$ taps using least mean square (LMS) algorithm trained on 50000 samples.

One important point to mention is that the CC we calculate in this work stands for Sy-based equalizers. The architecture of Sy-FNN/GRU/CNN is aimed to equalize a single symbol at the output, while the Sa-FNN/GRU/CNN is applied in the time domain equalizing a single sample at the output of NN. To quantify the complexity of Sa-based NNs the total $CC_{NN}$ has to be further multiplied by the number of $sps$, which is 8 in the simulation, and 2 in the experimental scenario. At the same time, the complexity of match filtering is not accounted for in this case.
\section{Results and Discussions}
\subsection{Performance Comparison}
\begin{figure}
    \centering
   \captionsetup{justification=centering,margin=2cm}
    \includegraphics[width=0.45\textwidth]{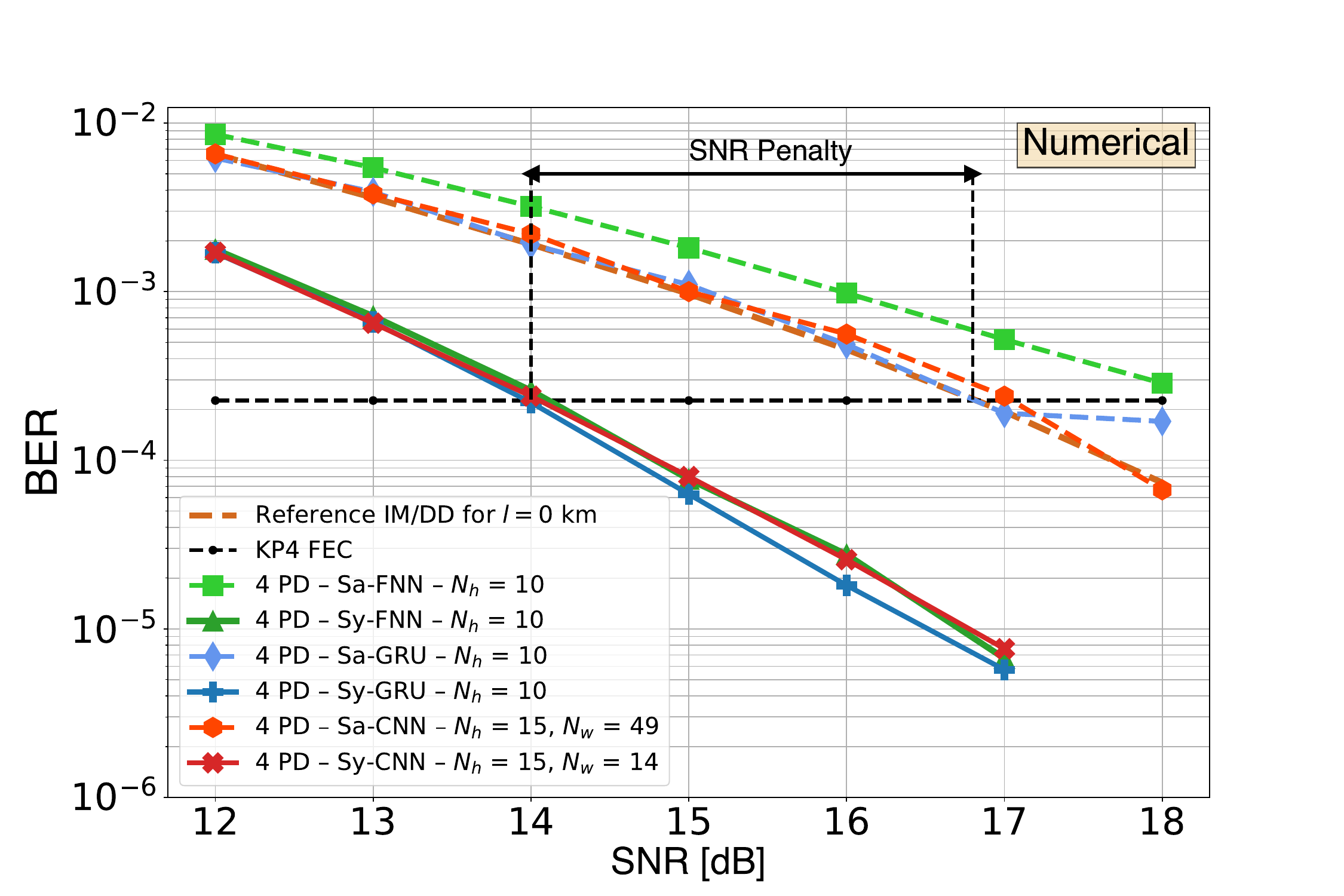}
    \caption{BER versus SNR at $l=74$ km transmission for all NN-based equalizers.}
    \label{fig:results0}
\end{figure}

To compare and evaluate the Sa-NN and Sy-NN equalizers in numerical simulations, the BER versus SNR performance for $l=74$ km transmission is shown in Fig.~\ref{fig:results0}. First, the back-to-back (B2B) transmission is referenced at KP4 forward error correction (FEC) threshold (BER~$=2.24 \times 10^{-4}$). Fig.~\ref{fig:results0} shows that the proposed Sy-NN approach outperforms the Sa-NN for all the types of NN equalizers. Because the Sy-based NN learns not only to compensate for the impairments but also to approximate the output OOK symbol in a regression way, optimizing it as a single DSP block improves transmission reach. Compared to Sy-NN, the Sa-NN is optimized disregarding the following RRC filter. Although both types of equalization approaches have a similar $K=3$ number of previous and future symbols in the input, the Sa-based NNs cannot properly compensate the ISI and equalize the samples into a correct symbol using a fixed post-processing RRC-based match filtering. Additionally, the Sy-based approach has a single output neuron which is trained to provide the symbol output between 0 and 1. However, for the Sa-based, the output neuron corresponds to a value in a broader range of a time domain input signal. The output neuron of a Sa-NN also tries to provide the values in between the sample-domain range, while the Sy-NN aims to output values only close to 0 and 1. Even though both methods have the same degrees of freedom at the input obtained from the neighboring samples, the Sa-GRU outperforms the Sa-FNN and Sa-CNN having a 0.8 dB KP4 SNR penalty. This improvement can be understood as GRU cells involve higher internal multiplicative complexity by memorizing previously calculated states, that are added to a short-term memory from the input. As for the Sy-based approach, all three equalizers Sy-FNN, Sy-GRU, and Sy-CNN show similar equalization performance improving the required SNR penalty at KP4 FEC threshold by around 2.8 dB compared to the un-equalized B2B reference. To quantify the impact of the NN-based equalizers on the transmission reach we calculate the SNR penalty at KP4 FEC threshold as a difference in dB between the transmission with equalization and the un-equalized reference IM/DD for $l=0$ km (Fig.~\ref{fig:results0} - SNR Penalty).
\begin{figure}
    \centering
   \captionsetup{justification=centering,margin=2cm}
    \includegraphics[width=0.45\textwidth]{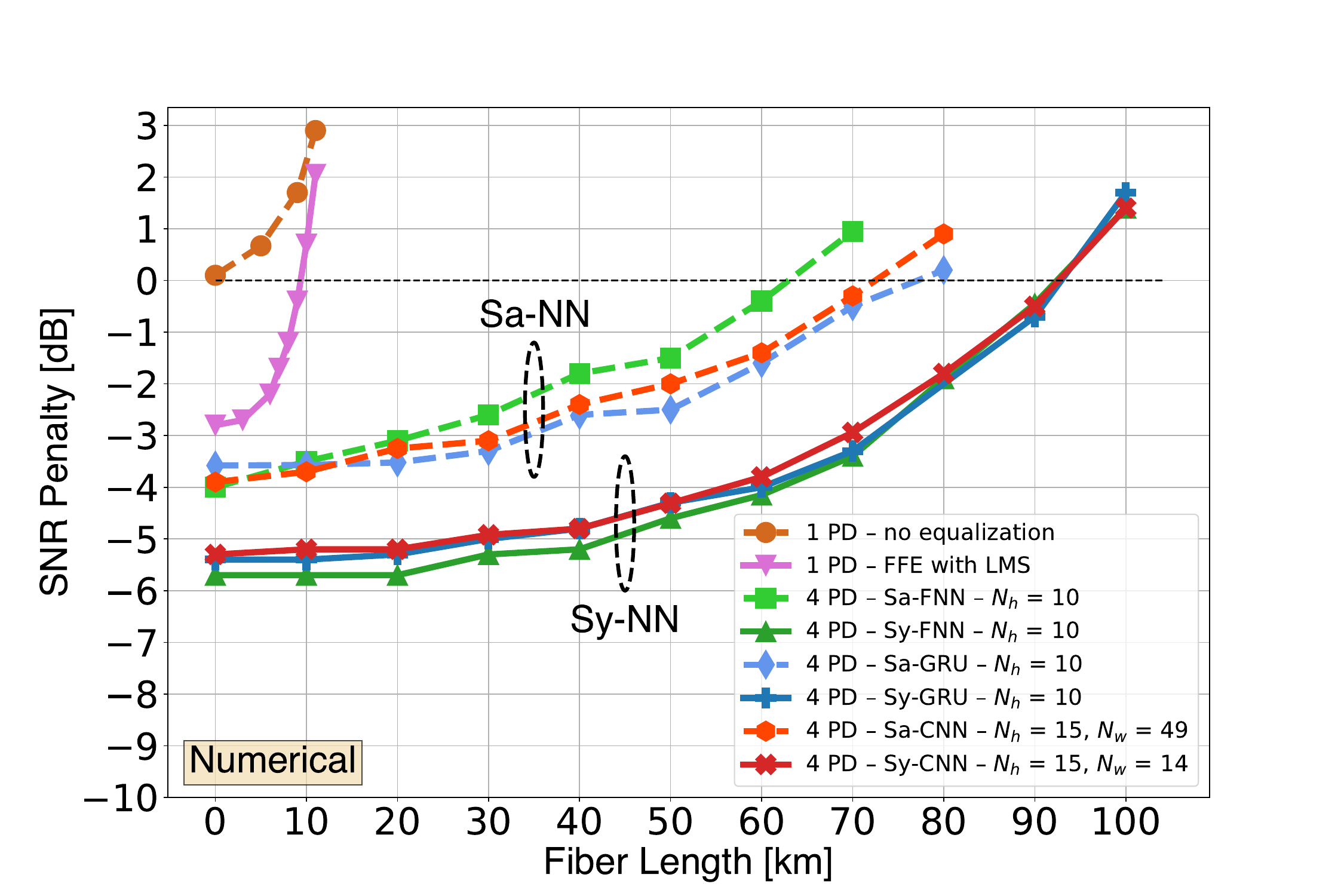}
    \caption{SNR penalty at the KP4 FEC threshold with respect to B2B un-equalized performance versus the fiber length.}
    \label{fig:results1}
\end{figure}

To further compare analyzed equalization approaches, Fig.~\ref{fig:results1} shows the KP4 FEC SNR penalty versus the transmission distance of the communication system. First, the single PD receiver with no equalization is plotted as a reference. To show the inability of compensation for a longer transmission distance of a conventional feedforward equalizer, the FFE with LMS is also applied~\cite{4476459}. The Sa-NN approaches show the robust KP4 FEC transmission reach of up to 75-km outperforming the FFE equalizer. It can be noticed that the simple Sa-FNN shows similar to Sa-GRU and Sa-CNN performance up to 20-km transmission. However, the strong impact of the ISI combined with insufficient input memory leads to more errors in the output for the Sa-FNN at longer distances. Therefore, the Sa-FNN can reach only up to 63 km transmission. In contrast to that, the Sa-CNN and Sa-GRU have more internal complexity, which slightly increases the equalization capacity and improves the transmission reach to 71-km and 74-km accordingly. As for the Sy-NN, the proposed Sy-based equalization approach outperforms the Sa-based one by 2 dB on average for all transmissions. Due to the sufficient amount of short-term input memory and simpler feed-forward structure, the Sy-FNN performs better for shorter distances of up to 50 km. However, all three Sy-FNN, Sy-GRU, and Sy-CNN show identical performance and increase the transmission reach without penalty up to 93 km. It shows that the short-term memory introduced at the input plays a crucial role in the equalization capacity of a symbol output Sy-NN structure.

\begin{center}
\begin{table}
\caption{NN complexity realization for numerical scenario \label{tab:table3}}
\centering
\scalebox{0.86}{
\begin{tabular}{ |c||c|c||c|c||c|c||c|c||c|c| }
\hline
 & \multicolumn{2}{c||}{$10^2$} & \multicolumn{2}{c||}{$2\times10^2$} & \multicolumn{2}{c||}{$5\times10^2$} & \multicolumn{2}{c||}{$10\times10^2$} & \multicolumn{2}{c|}{$15\times10^2$} \\
 \hline
 & $M$ & $N_h$ & $M$ & $N_h$ & $M$ & $N_h$ & $M$ & $N_h$ & $M$ & $N_{h}$ \\
\hline
 Sy-FNN & 14 & 1 & 14 & 2 & 14 & 5 & 14 & 9 & 14 & 14 \\
\hline
 Sa-FNN & - & - & - & - & - & - & - & - & 49 & 1 \\
\hline
 Sy-GRU & 6 & 1 & 14 & 1 & 14 & 2 & 14 & 3 & 14 & 4 \\
\hline
 Sa-GRU & - & - & - & - & - & - & - & - & - & - \\
\hline
 Sy-CNN & 14 & 2 & 14 & 4 & 14 & 10 & 14 & 18 & 14 & 27 \\
\hline
 Sa-CNN & - & - & - & - & - & - & - & - & 49 & 1 \\
\hline
\end{tabular}}
\end{table}
\end{center}

\begin{center}
\begin{table}
\caption{NN complexity realization for experimental scenario \label{tab:table2}}
\centering
\scalebox{0.86}{
\begin{tabular}{ |c||c|c||c|c||c|c||c|c||c|c| }
\hline
 & \multicolumn{2}{c||}{$10^2$} & \multicolumn{2}{c||}{$2\times10^2$} & \multicolumn{2}{c||}{$5\times10^2$} & \multicolumn{2}{c||}{$10\times10^2$} & \multicolumn{2}{c|}{$15\times10^2$} \\
 \hline
 & $M$ & $N_h$ & $M$ & $N_h$ & $M$ & $N_h$ & $M$ & $N_h$ & $M$ & $N_{h}$ \\
\hline
 Sy-FNN & 14 & 1 & 14 & 2 & 14 & 5 & 14 & 9 & 14 & 14 \\
\hline
 Sa-FNN & 13 & 2 & 13 & 4 & 13 & 10 & 13 & 18 & 13 & 27\\
\hline
 Sy-GRU & 6 & 1 & 14 & 1 & 14 & 2 & 14 & 3 & 14 & 4 \\
\hline
 Sa-GRU & 5 & 1 & 9 & 1 & 13 & 1 & 13 & 2 & 13 & 3 \\
\hline
 Sy-CNN & 14 & 2 & 14 & 4 & 14 & 10 & 14 & 18 & 14 & 27 \\
\hline
 Sa-CNN & 13 & 1 & 13 & 2 & 13 & 5 & 13 & 9 & 13 & 14 \\
\hline
\end{tabular}}
\end{table}
\end{center}
\vspace{-42pt}
\subsection{Computational Complexity Comparison}
\begin{figure}[!t]
    \centering
   \captionsetup{justification=centering,margin=2cm}
    \includegraphics[width=0.45\textwidth]{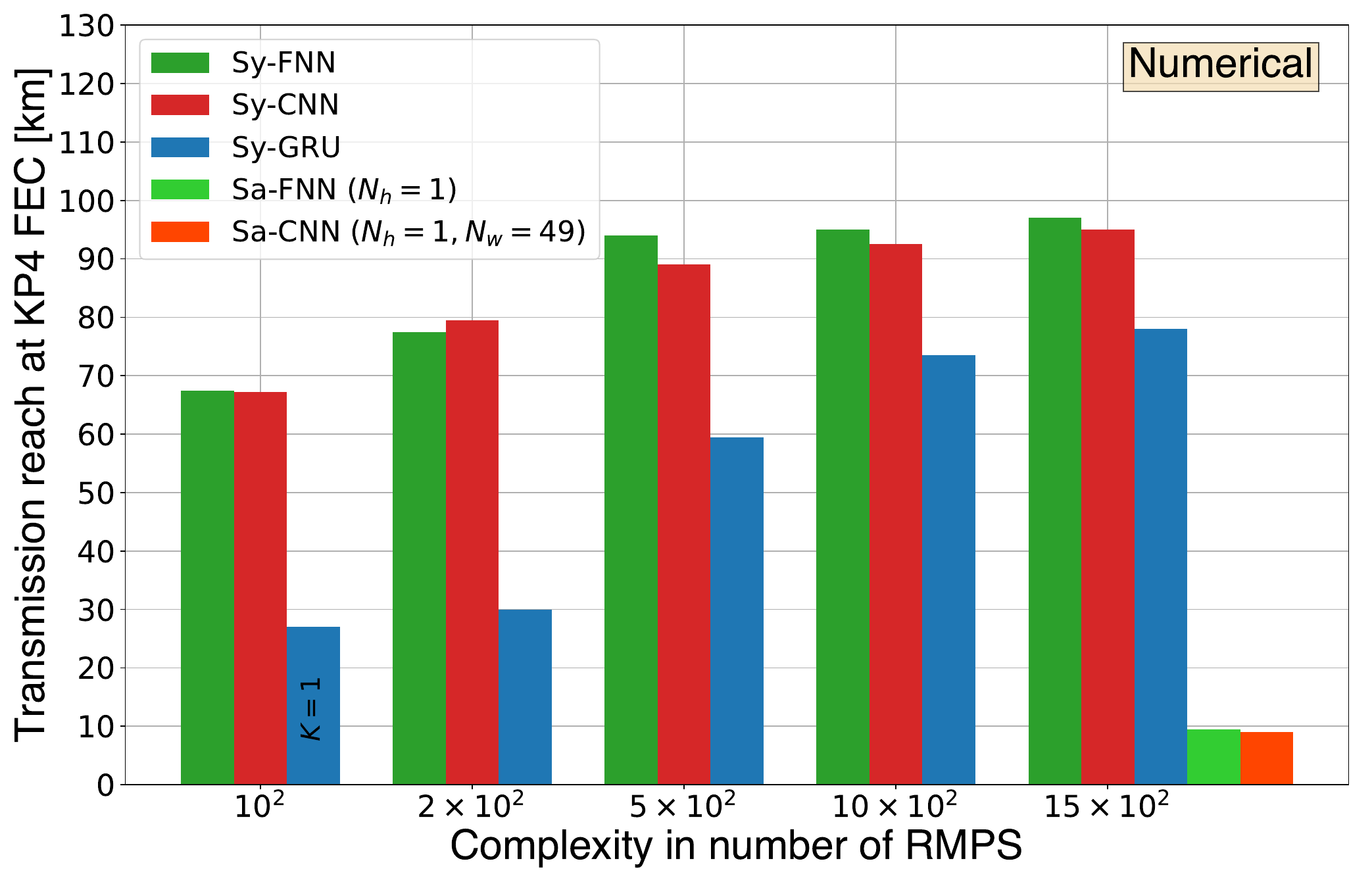}
    \caption{Computational complexity in RMPS for numerical analysis.}
    \label{fig:results2}
\end{figure}

Computational complexity in terms of RMPS of NN-based equalizers is an essential aspect when considering simple IM/DD systems. To evaluate the proposed NN-based equalizers from the complexity angle, we limit the available number of RMPS and define the range from 100 to 1500 multiplications. To vary the number of RMPS we change the number of hidden units in FNN and GRU, and filters in CNN architectures. The corresponding optimized hyperparameters used for both Sa-NN and Sy-NN architectures are shown in Table~\ref{tab:table3}. Additionally, we do not include the complexity of the matching filter for Sa-based equalizers which will also increase the total CC but with a constant, architecture-independent offset. Therefore, our main focus here is to investigate the Sy-NN architectures and evaluate their performance at different complexity levels. Fig.~\ref{fig:results2} shows the achievable transmission reach without penalty for restricted complexity levels in the number of RMPS. It is shown, that the Sy-FNN and Sy-CNN can reach similar transmission performance for $10^2$, and $2\times10^2$ RMPS. However, starting from $5\times10^2$ RMPS, the Sy-FNN reaches its equalization capacity for 93 km transmission and keeps similar transmission reach up to $15\times10^2$ RMPS. While Sy-CNN reaches similar to Sy-FNN equalization performance only at $15\times10^2$ RMPS. As for the Sy-GRU architecture, restricting the number of RMPS to a couple of hundred RMPS limits the equalization capacity of GRU cells significantly decreasing the transmission reach. Due to the complex gated structure, to define the Sy-GRU with $10^2$ RMPS, we had to decrease the input memory $M$ by setting the number $K$ in Eq.~\ref{eq:1} to $K=1$~symbol, which led to a decrease in equalization performance. For the rest of the GRU complexity levels the $K$ is kept fixed equal to 3. It can be seen, that the Sy-GRU simply does not have enough multiplicative capability to compensate memory-related ISI at longer distances. Additionally, for the CC comparison, the Fig.~\ref{fig:results2} also reports the minimum possible Sa-FNN and Sa-CNN structures, demonstrating the inability of the equalizers to properly compensate for the impairments within limited complexity.\

\begin{figure}[t]
    \centering
   \captionsetup{justification=centering,margin=2cm}
    \includegraphics[width=0.45\textwidth]{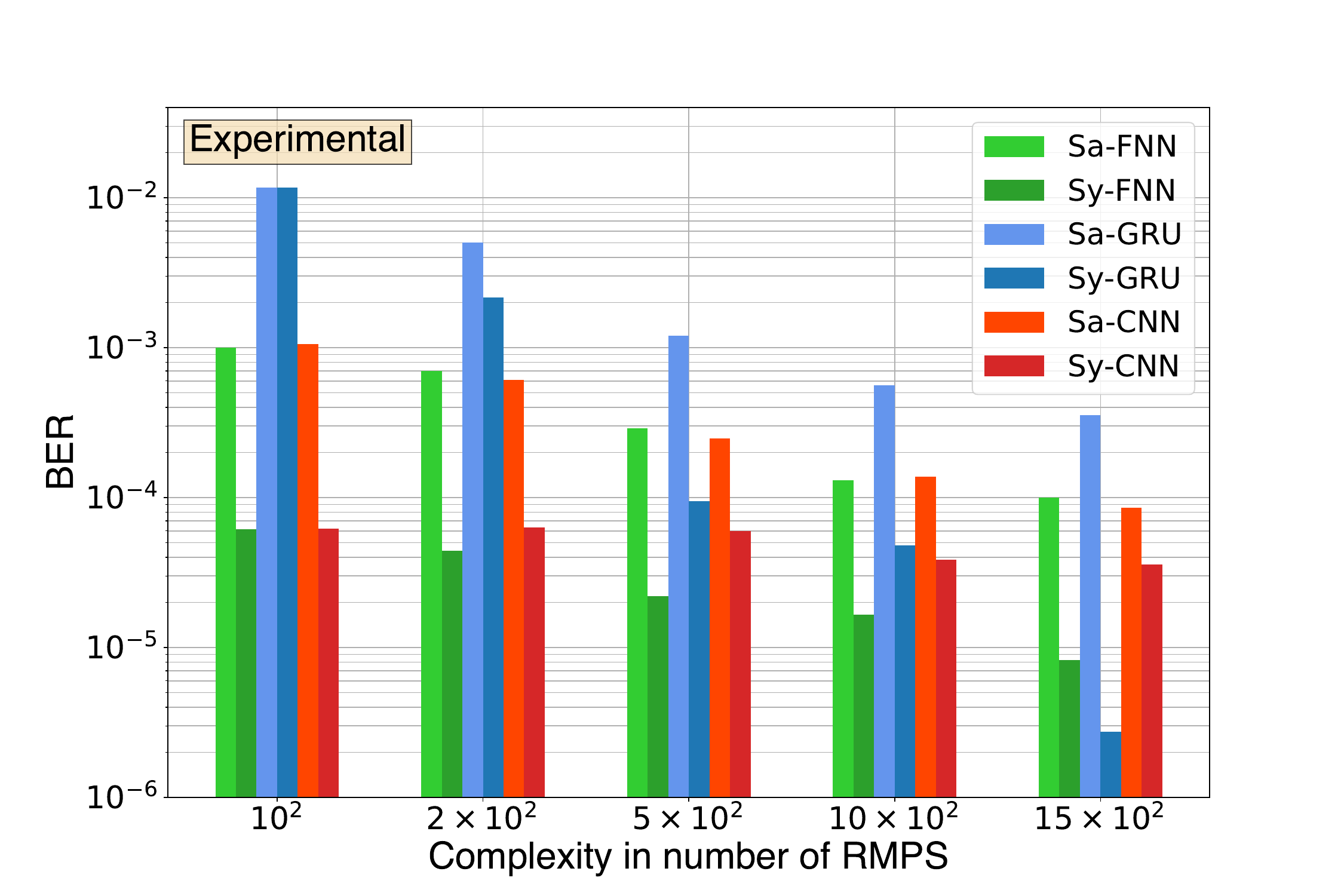}
    \caption{BER for experimental validation of NN with restricted RMPS at $l=74$ km.}
    \label{fig:results3}
\end{figure}

To validate the numerical results of the proposed approach, we experimentally compare the equalization performance of Sy and Sa-based equalizers applied for 32-GBd 74-km transmission in SSMF. The BER versus the restricted complexity number is shown in Fig.~\ref{fig:results3}. It is important to highlight that in the experimental setup, the signal is upsampled to $sps=2$. Therefore, the CC of Sa-NN will be adapted accordingly. The memory $M$ and the number of hidden units for corresponding CC levels are summarized in Table~\ref{tab:table2}. It can be seen, that optimizing the Sy-NN as a single DSP block outperforms the Sa-based equalization for all the investigated NN architectures. The Sy-FNN, particularly, shows the lowest BER compared to Sy-GRU and Sy-CNN for the CC levels up to $10\times10^2$. However, at the $15\times10^2$ number of CC, the Sy-GRU outperforms the Sy-FNN due to higher recurrent memory that allows for storing essential information about the accumulated impairments. It can be seen, that for the CC level of $10^2$ the Sa-GRU and Sy-GRU perform equally poorly, because a single hidden GRU cell is unable to capture the dynamics of the transmission impairments. However, increasing the number of hidden GRU cells along with the input memory leads to higher Sy-GRU and Sa-GRU equalization capacity and lower BER. For the  Sy-CNN, the combination of filters is unable to compensate for the memory-induced impairments. Similarly to numerical results, the proposed Sy-NN equalization outperforms the Sa-NN in terms of BER performance for all levels of CC. This leads to the conclusion that when the CC is around a couple of hundred multiplications the Sy-FNN equalizer is a primary candidate for a simple IM/DD setup. However, increasing both input memory and hidden recurrent units can increase the equalization performance at the expense of higher CC.\

\section{Conclusions}
In this work, we demonstrate that designing a samples-to-symbol neural network (NN)-based equalizer as (Sy-NN) outperforms a samples-to-sample neural network (Sa-NN) for compensating the impairments in 32 GBd on-off keying intensity-modulated and directly-detected transmission in both numerical and experimental scenarios. Numerical simulations show that the Sy-NNs can efficiently transform 4 slices of upsampled pulse-shaped signals into a single symbols output increasing the transmission reach up to 93 km. This design minimizes the computational complexity (CC) of the internal NN, eliminating the need for external digital signal processing (DSP) blocks, like matched filters. Additionally, experimental transmission over 74 km validates the numerical results, showing an improvement of an order of magnitude for the Sy-FNN over the Sa-FNN. Comparing the CC of the Sy-based equalizers, we show that using the FNN delivers high equalization performance while keeping the complexity at a couple of hundred real multiplications per equalized symbol.
 
\section*{Acknowledgments}
This work was financially supported by the Villum Fonden’s YIP OPTIC-AI project (grant no 29344) and ERC CoG FRECOM (grant no. 771878).

\bibliographystyle{IEEEtran}
\bibliography{references}

\vfill
\end{document}